\documentclass[a4paper,11pt]{article}

  \usepackage{
    a4wide,
    amsmath,
    amsfonts,
    bm,
    braket,
    cite,
    comment,
    enumerate,
    here,
    mathtools,
    }
  \usepackage[italicdiff]{physics}
  \usepackage{color,graphicx}
  \usepackage[labelsep=quad]{caption}
  \usepackage[labelsep=quad]{subcaption}
  \usepackage{authblk}
  \usepackage{tikz}

\usepackage[normalem]{ulem}

  \makeatletter
\@addtoreset{equation}{section}
\def\theequation{\thesection.\arabic{equation}}
\makeatother

\newcommand{\figref}[1]{Fig.~\ref{#1}}

\newcommand{\e}[1]{\mathrm{e}^{#1}}

\newcommand{\veps}{\varepsilon}

\newcommand{\fcn}[2]{{{#1}\pqty{#2}}}
\newcommand{\fcl}[2]{{{#1}\bqty{#2}}}
\newcommand{\nt}{\notag\\}

\definecolor{DarkMagenta}{rgb}{0.54,0,0.54} 

\definecolor{DarkBlue}{rgb}{0,0,0.7} 

\definecolor{DarkRed}{rgb}{0.65,0,0}

\begin{document}

\title{
Gravothermal catastrophe and critical dimension in a $D$-dimensional asymptotically AdS spacetime
}

\author{Hiroki Asami\footnote{Email:\href{mailto:asami.hiroki.b3@s.mail.nagoya-u.ac.jp}{asami.hiroki.b3@s.mail.nagoya-u.ac.jp}}}
\author{Chul-Moon Yoo\footnote{Email:\href{mailto:yoo@gravity.phys.nagoya-u.ac.jp}{yoo@gravity.phys.nagoya-u.ac.jp}}}
\affil{Division of Particle and Astrophysical Science, Graduate School of Science, Nagoya University, Nagoya 464-8602, Japan}
\setcounter{Maxaffil}{0}
\renewcommand\Affilfont{\itshape\small}
\date{}

\maketitle

\begin{abstract}
We investigate the structure and stability of the thermal equilibrium states of a spherically symmetric self-gravitating system in 
a $D$--dimensional asymptotically Anti-de Sitter(AdS) spacetime.
The system satisfies the Einstein-Vlasov equations with a negative cosmological constant.
Due to the confined structure of the AdS potential, we can construct thermal equilibrium states 
without any artificial wall in the asymptotically AdS spacetime. 
Accordingly, the AdS radius can be regarded as the typical size of the system.
Then the system can be characterized by the gravothermal energy and AdS radius normalized by the total particle number. 
We investigate the catastrophic instability of the system
in a $D$--dimensional spacetime by using the turning point method.
As a result, we find that the curve has a double spiral structure for $4\le D\le 10$ while it 
does not have any spiral structures for $D\ge11$ as  in the asymptotically flat case confined by an adiabatic wall. 
Irrespective of the existence of the spiral structure, 
there exist upper and lower bounds for the value of the gravothermal energy. 
This fact indicates that there is no thermal equilibrium solution outside the allowed region of the gravothermal energy.
This property is also similar to the asymptotically flat case.
\end{abstract}



\section{Introduction}
Since the seminal work of Antonov~\cite{antonov1962}, the thermodynamical and statistical properties of 
the self-gravitating many-particle systems have attracted much attention. 
Antonov investigated thermal equilibrium states of the self-gravitating system surrounded by a spherical adiabatic wall of radius $R$ in the microcanonical ensemble
\footnote{In the microcanonical ensemble, the thermal relaxation leads the extremum Boltzmann-Gibbs entropy state 
with fixed energy and particle number in the wall. The adiabatic wall plays a role in preventing the particles from diffusion.}.
In Ref~\cite{antonov1962}, it has been shown that the Boltzmann-Gibbs entropy has no global maximum but may have a local maximum.
Furthermore, it has been also shown that the second variation of the entropy with respect to the distribution function can be positive if the density contrast, 
which is given by the ratio between the mass density at the center and the edge, is greater than 709.
This result indicates that the density contrast must not exceed 709 for the thermal equilibrium state to be stable.
Lynden-Bell and Wood~\cite{BellWood1968} confirmed and extended the results of~\cite{antonov1962} in accordance with Poincare's stability criterion~\cite{Poincare1885}.
Ref.~\cite{Poincare1885} argued that a series of equilibria can increase the number of unstable modes only if it passes a turning point~\cite{katz1978,Katz1979,Sorkin1981,Sorkin1982}.
They found the existence of the minimum value of the gravothermal energy associated with the turning point which is identical to the critical point found by Antonov in the microcanonical ensemble.

This catastrophic instability of the self-gravitating systems is called the gravothermal catastrophe, 
which is a consequence of the negative specific heat of gravity.
The negative specific heat makes the core of the system be heated when it transfers energy to the surrounding halo.
Once it begins, the energy current carried by particles continues to be transferred from the core to the halo, and fragmentation will occur. 
In the Newtonian case, this catastrophic instability does not appear if the thermal energy dominates over the gravitational potential energy and the confinement is sufficiently effective.
This is because the adiabatic wall puts external pressure on the system and makes the specific heat positive.
This situation corresponds to the case where the density contrast is smaller than 709, 
and it can be realized by increasing the thermal energy under fixed values of the particle number and the size of the confining wall.

To perform the relativistic extension, 
we need to take into account the contribution of the thermal energy to the total mass. 
Thus the increasing thermal energy raises the compactness of the system defined by the total mass divided by the radius.
In general relativity, the system cannot avoid gravitational collapse if it is sufficiently compact~\cite{Roupas2014,Roupas:2014sda,Roupas_2015,Roupas2018,Alberti:2019xaj,Chavanis2020,SWZ1981,Chavanis:2007kn}. 
As a result, the series of equilibria has a double spiral structure, and the two catastrophic instabilities appear in association with two turning points.
For a relativistic perfect fluid with a linear equation of state, the dependence of the stability on the number of spacetime dimensions has been studied in Ref.~\cite{Chavanis:2007kn}.
In an asymptotically flat spacetime, Chavanis concluded that the system has the critical dimension $D_\ast$ such that the system has a double spiral structure for $D<D_\ast$ and not for $D \ge D_\ast$.
In an asymptotically AdS spacetime, motivated by the Hawking-Page phase transition, 
Refs.~\cite{Vaganov:2007at,Hammersley:2007ahw} investigated the stability of self-gravitating radiation and obtained the same result,
i.e., the double spiral structure vanishes above the critical dimension $D_\ast$. 

Asymptotically AdS spacetimes have also been attractive in the context of the AdS/CFT correspondence~\cite{Maldacena_1999,Gubser_1998,witten1998anti} and the gravitational turbulent instability~\cite{Bizo__2011}. 
Since an asymptotically AdS spacetime has a confined structure, perturbations of matter fields in the spacetime may not dissipate to infinity. 
Due to the nonlinearity of the field equations, they interact with each other and may form black holes even if the perturbations are arbitrarily small.
Despite many works on the turbulent instability~\cite{Dias:2012tq,Maliborski:2013jca,Buchel:2013uba,Balasubramanian:2014cja,Balasubramanian:2015uua,Bizon:2015pfa,Dimitrakopoulos:2015pwa,Green:2015dsa,Garfinkle:2011hm,Jalmuzna:2011qw,Bizon:2017yrh,Craps:2014vaa,Craps:2014jwa,Evnin:2021buq}, 
the final fate has not been clarified yet because the time evolution and final states are dependent on symmetries, dimensions and boundary conditions of the spacetime~\cite{Choptuik_2018,Masachs_2019}.
In addition to the case of matter fields, the stability of the asymptotically AdS Einstein-Vlasov system is also investigated in Ref.~\cite{Moschidis:2017llu, Asami_2021} (see also~\cite{Gunther:2020mvb,Moschidis:2017lcr})
\footnote{
  Besides, the stability of the massless Einstein--Vlasov system with an inner mirror has been investigated in Ref.~\cite{Moschidis:2017lcr} and 
  Ref.~\cite{Gunther:2020mvb} analyzed the stability of steady states of the spherically symmetric Einstein-Vlasov system in a flat spacetime.}.
However, the conditions of black formation 
and the final fate are still not completely clear~\cite{Choptuik_2018,Masachs_2019}. 
Since the self-gravitating many-particle system might be regarded as a macroscopic model for the excitations in an asymptotically AdS spacetime,
the analysis in our paper might be helpful to guess the final states of asymptotically AdS spacetimes in general dimensions.

This paper is organized as follows.
In section \ref{sec:eqs}, we derive the basic equations for the thermal equilibrium states.
The details of the derivation are shown in \ref{sec:app_der} and calculations for physical quantities are shown in \ref{sec:app_int_mom}.
We present the results in section \ref{sec:results}.
In section \ref{sec:profile}, we show the parameter dependence of the radial profile of the system.
After the brief review on the turning point method in section \ref{sec:method},
we show the main results of the stability analysis in section \ref{sec:structure}-\ref{sec:high_dim}. 
Section \ref{sec:conclusion} is devoted to a summary and conclusion.

Throughout this paper, we use the geometrized units in which both the speed of light and gravitational constant in $D$--dimension are unity, $c=G_D=1$.

\section{Field equations and thermodynamical quantities}
\label{sec:eqs}
\subsection{Physical quantities of the self-gravitating system}
  We consider a self-gravitating system of massive point particles in a $D$--dimensional spacetime.
  For simplicity, setting the rest mass of each particle to unity, we can identify the number of particles with the total rest mass. 
  Focusing on a static and spherically symmetric spacetime, 
  the metric is given by 
  \begin{align}
    g_{\mu\nu}\dd{x^\mu}\dd{x^\nu} = -\e{2\mu(r)}\dd{t^2}+\e{2\nu(r)}\dd{r^2}+r^2\dd\Omega_{D-2}^2,
    \label{eq:metric}
  \end{align}
  where 
  \begin{align}
    \dd\Omega_{D-2}^2
    =\dd\theta_1^2+\sin^2\theta_1\dd\theta_2^2+\sin^2\theta_1\sin^2\theta_2\dd\theta_3^2
      +\cdots
      +\sin^2\theta_1\sin^2\theta_2\cdots\sin^2\theta_{D-3}\dd\theta_{D-2}^2.
  \end{align}
  Introducing the one-particle distribution function $f(x^\mu,p^i)$ in the mean field approximation,
  the energy-momentum tensor of this system is written as
  \begin{align}
    T_{\mu\nu}:=\int dV_p p_\mu p_\nu f(x^\mu,p^i),
    \label{eq:def_emt}
  \end{align}
  where $\dd{V_p}$ is the invariant volume element in the momentum space.
  The on-shell condition $p^2+1=0$ leads to
  \begin{align}
    dV_p 
    &= 2\sqrt{-g}\delta(p^2+1)\theta(\veps)\dd[D]{p} \notag \\
    &= \frac{\sqrt{-g}}{\veps} \dd{p}^r\wedge\dd{p}^{\theta_1}\wedge\cdots\wedge\dd{p}^{\theta_{D-2}},
  \end{align}
  where $\delta$ and $\theta$ are the delta function and the Heaviside's step function, respectively, and $\veps := -p_t$ is the energy of the particle.
  The energy density and the pressure are defined by $\rho(r):=-T_t^{\ t}$ and $p(r):=T_r^{\ r}$, respectively.
  We introduce the quasi--local mass $M(r)$ as
  \begin{align}
    M(r) := S_{D-1}\int_0^r \dd{u}u^{D-2}\rho(u),
  \end{align}
  where $S_{D-1}$ is the surface area of the $D$-sphere:
  \begin{align}
    S_{D-1}:=\frac{2\pi^{\frac{D-1}{2}}}{\Gamma(\frac{D-1}{2})},
  \end{align}
  with $\Gamma$ being the Gamma function. 
  For a given function of the distribution function as $\mathcal{F} = \fcn{\mathcal{F}}{f}$, 
  the current
  \begin{align}
    \fcl{F^\mu}{\fcn{\mathcal{F}}{f}} = \int \dd{V_p}p^\mu\fcn{\mathcal{F}}{f}
    \label{eq:def_current}
  \end{align}
  carries the density $\fcn{\mathcal{F}}{f}$ and we can define the charge $F$ as
  \begin{align}
    \fcl{F}{\fcn{\mathcal{F}}{f}} := \int_\Sigma \dd{\Sigma_\mu}F^\mu, 
    \label{eq:def_charge}
  \end{align}
  where $\Sigma$ is a constant time hyper-surface.
  By substituting $\fcn{\mathcal{F}}{f} = f$, we can define the number of particles within the radius $r$ as 
  \begin{align}
    N(r):=\eval{\fcl{F}{\mathcal{F}}}_{\fcn{\mathcal{F}}{f}=f} = \int_0^r\dd{u}u^{D-2} n(u),
  \end{align}
  where $n(r):=\e{\mu+\nu}N^t$ with $N^\mu := \eval{\fcl{F^\mu}{\mathcal{F}}}_{\fcn{\mathcal{F}}{f}=f}$.
  We also define the Boltzmann-Gibbs (BG) entropy as 
  \begin{align}
    S(r):=\eval{\fcl{F}{\mathcal{F}}}_{\fcn{\mathcal{F}}{f}=-f\pqty{\log f-1}} = \int_0^r\dd{u}u^{D-2} s(u),
  \end{align}
  where $s(r):=\e{\mu+\nu}S^t$ with $S^\mu := \eval{\fcl{F^\mu}{\mathcal{F}}}_{\fcn{\mathcal{F}}{f}=-f\pqty{\log f-1}}$.

\subsection{Einstein's equations}
  The Einstein's equations with a negative cosmological constant $G_{\mu\nu}+\Lambda g_{\mu\nu} = 8\pi T_{\mu\nu}$ reduce to the following equations:
  \begin{subequations}
    \begin{align}
      -16\pi r^2\rho(r)\e{2\nu} &= (D-3)(D-2)+\e{2\nu}\pqty{-(D-3)(D-2)+2\Lambda r^2}-2(D-2)r\nu^\prime,
      \label{eq:ein_1}\\
      16\pi r^2p(r)\e{2\nu}     &= (D-3)(D-2)+\e{2\nu}\pqty{-(D-3)(D-2)+2\Lambda r^2}+2(D-2)r\mu^\prime.
      \label{eq:ein_2}
    \end{align}
  \end{subequations}
  Integrating Eq.~\eqref{eq:ein_1} with the central regularity: $\nu(0)=0$, we obtain
  \begin{align}
    \e{-2\nu} = 1-\frac{kM(r)}{r^{D-3}}+\frac{r^2}{L^2}\qc k := \frac{16\pi}{(D-2)S_{D-1}}, 
    \label{eq:ein_3}
  \end{align}
  where $L$ is the AdS radius defined as
  \begin{align}
    L := \sqrt{-\frac{(D-1)(D-2)}{2\Lambda}}.
  \end{align}
  By using Eq.~\eqref{eq:ein_3}, Eq.~\eqref{eq:ein_2} becomes 
  \begin{align}
    \mu^\prime = \frac{\frac{k(D-3)M(r)}{2r^{D-3}}+\frac{r^2}{L^2}+\frac{8\pi r^2p(r)}{D-2}}{r\pqty{1-\frac{kM(r)}{r^{D-3}}+\frac{r^2}{L^2}}}. 
    \label{eq:ein_main}
  \end{align}

  If the central values of the energy density $\rho_c$ and the pressure $p_c$ do not vanish, 
  we can normalize the physical quantities by using them together with $\mu_c:=\mu(0)$ .
  Introducing the typical length scale $\ell := \pqty{S_{D-1}\rho_c}^{-1/2}$ and 
  defining $x := r/\ell$, $y := \mu-\mu_c$, $\tilde{\rho} := \rho/\rho_c$, $\tilde{n} := n/n_c$ and $\tilde{p} := p/p_c$,
  we obtain the mass and the particle number as
  \begin{align}
    M(r) = \ell^{D-3}\int_0^x \dd{u}u^{D-2}\tilde{\rho}(u)\qc N(r) = \ell^{D-3}\frac{n_c}{\rho_c}\int_0^x \dd{u}u^{D-2}\tilde{n}(u).
  \end{align}
  Then introducing $\tilde{M}(x) := M(r)/\ell^{D-3}$ and $\tilde{N}(x) := N(r)/\ell^{D-3}$, we can rewrite Eq.~\eqref{eq:ein_main} as
  \begin{align}
    y^\prime = \frac{\frac{k(D-3)\tilde{M}(x)}{2x^{D-3}}+\frac{x^2}{\lambda^2}+\frac{8\pi w_c x^2 \tilde{p}(x)}{(D-2)S_{D-1}}}{x\pqty{1-\frac{k\tilde{M}(x)}{x^{D-3}}+\frac{x^2}{\lambda^2}}},
    \label{eq:ein_norm}
  \end{align}
  where $\lambda := L/\ell$ and $w_c := p_c/\rho_c$.
  By numerically solving Eq.~\eqref{eq:ein_norm} together with 
  \begin{align}
    \dv{\tilde{M}}{x}=x^{D-2}\tilde{\rho},
    \label{eq:ein_mass}
  \end{align}
  we can get a solution.
  The boundary conditions are given by $\fcn{y}{0} = \fcn{y^\prime}{0}=\fcn{\tilde{M}}{0}=0$ due to the definition of $\fcn{y}{x}$ and $\fcn{\tilde{M}}{x}$.
  We note that Eq.~\eqref{eq:ein_norm} is valid for $D \ge 3$, and the $D=3$ case must be independently treated.
  We focus on the $D \ge 4$ cases in this paper.

\subsection{Thermal equilibrium states}
\label{sec:TES}
  In the isolated system, the thermal equilibrium states are given as extremal entropy states 
  with a fixed total mass and total particle number.
  Introducing the positive numbers $\alpha$ and $\beta$ as the Lagrange multipliers, 
  the condition, which is called the Gibbs relation, is written as
  \begin{align}
    \var{S}+\alpha\var{N}-\beta\var{M}=0,
    \label{eq:extremal}
  \end{align}
  where $S:=\lim_{r\to\infty}\fcn{S}{r},\ N:=\lim_{r\to\infty}\fcn{N}{r}$ and $M:=\lim_{r\to\infty}\fcn{M}{r}$ are the total entropy, the total particle number and the total mass, respectively.
  As is shown in \ref{sec:app_der}, 
  this condition leads to the Maxwell-J\"{u}ttner (MJ) distribution function $f=\exp\pqty{\alpha-\beta\veps}$ in the $D$--dimensional static and spherically symmetric spacetime.
  As is shown in \ref{sec:app_int_mom}, for MJ distribution, performing the integral over the momentum space, 
  we obtain the energy density, the pressure and the number density as follows:
  \begin{subequations}
    \begin{align}
      \rho(r)
      &= 2\e{\alpha}\pqty{\frac{2\pi}{z}}^{\frac{D}{2}-1}\pqty{\frac{D-1}{z}K_{\frac{D}{2}}(z)+K_{\frac{D}{2}-1}(z)}, \\
      p(r)
      &= \frac{\e{\alpha}}{\pi}\pqty{\frac{2\pi}{z}}^{\frac{D}{2}}K_{\frac{D}{2}}(z), \\
      n(r)
      &= 2\e{\alpha+\nu}\pqty{\frac{2\pi}{z}}^{\frac{D}{2}-1}K_{\frac{D}{2}}(z),
    \end{align}
  \end{subequations}
  where $z=\beta\e{\mu}\eqqcolon\beta\veps_0$.
  By using the normalized variables, they are rewritten as
  \begin{subequations}
    \begin{align}
      \tilde{\rho}(x)
      &=\frac{(D-1)K_{\frac{D}{2}}(z)\e{-y}+z_cK_{\frac{D}{2}-1}(z)}{(D-1)K_{\frac{D}{2}}(z_c)+z_cK_{\frac{D}{2}-1}(z_c)}\e{-\pqty{\frac{D}{2}-1}y}, \\
      \tilde{p}(x)
      &=\frac{K_{\frac{D}{2}}(z)}{K_{\frac{D}{2}}(z_c)}\e{-\frac{D}{2}y}, \\
      \tilde{n}(x)
      &=\frac{K_{\frac{D}{2}}(z)}{K_{\frac{D}{2}}(z_c)}\pqty{1-\frac{k\tilde{M}(r)}{x^{D-3}}+\frac{x^2}{\lambda^2}}^{-\frac{1}{2}}\e{-\pqty{\frac{D}{2}-1}y}
    \end{align}
  \end{subequations}
  and 
  \begin{align}
    w_c
    &= \frac{K_{\frac{D}{2}}(z_c)}{(D-1)K_{\frac{D}{2}}(z_c)+z_cK_{\frac{D}{2}-1}(z_c)},
  \end{align}
  where $z_c := \beta\e{\mu_c}$, which is the inverse temperature including the effect of the lapse function at the center.
  Therefore the parameter $z_c$ corresponds to the red-shft factor at the center and it characterizes the relativistic effect of the system.
  We can also rewrite $\tilde{M}(x)$ and $\tilde{N}(x)$ as
  \begin{align}
    \tilde{M}(x) = \int_0^x\dd{z}z^{D-2}\tilde{\rho}(z)\qc
    \tilde{N}(x) = z_c w_c \int_0^x \dd{z}z^{D-2}\tilde{n}(z)
  \end{align}
  because $n_c/\rho_c = n_c/p_c\cdot p_c/\rho_c = z_c w_c$ is satisfied.
  Normalizing the physical quantities by the particle number, we obtain
  \begin{subequations}
    \begin{align}
      \tilde{E} &\coloneqq \frac{E}{N} = \frac{\int dx x^{D-2}\tilde{\rho}}{z_cw_c\int dx x^{D-2}\tilde{n}}-1, \\
      \tilde{L}^{D-3} &\coloneqq \frac{L^{D-3}}{N} = \frac{\lambda^{D-3}}{z_cw_c\int dx x^{D-2}\tilde{n}}, \\
      \tilde{r}^{D-3} &\coloneqq \frac{r^{D-3}}{N} = \frac{x^{D-3}}{z_cw_c\int dx x^{D-2}\tilde{n}},
    \end{align}
    \label{phys_quat}
  \end{subequations}
  where $E := M-N$ is the energy except for the rest mass of the particles, which is called the gravothermal energy.

  \section{Results}
  \label{sec:results}
    \subsection{Parameter dependence of the profile}
    \label{sec:profile}
\begin{figure}
      \begin{subfigure}[]{0.5\linewidth}
        \centering
        \hspace{-20pt}
        \includegraphics[width = 8cm]{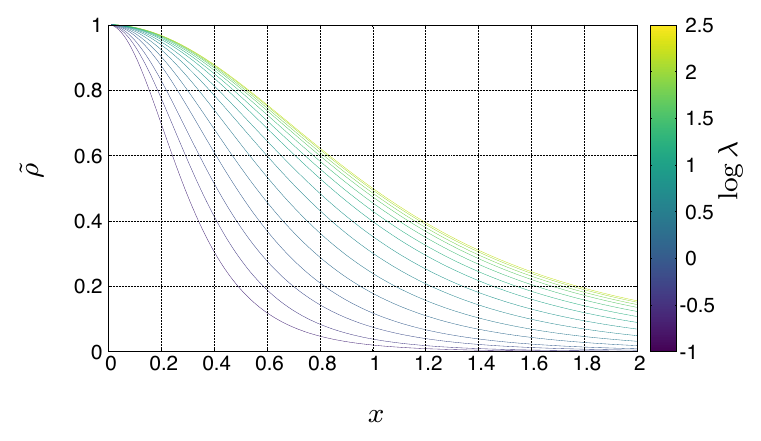}
        \subcaption{$\lambda$ dependence for $z_c = 1$.}
        \label{fig:dim5_lambda_dep}
      \end{subfigure}
      \begin{subfigure}[]{0.5\linewidth}
        \centering
        \includegraphics[width = 8cm]{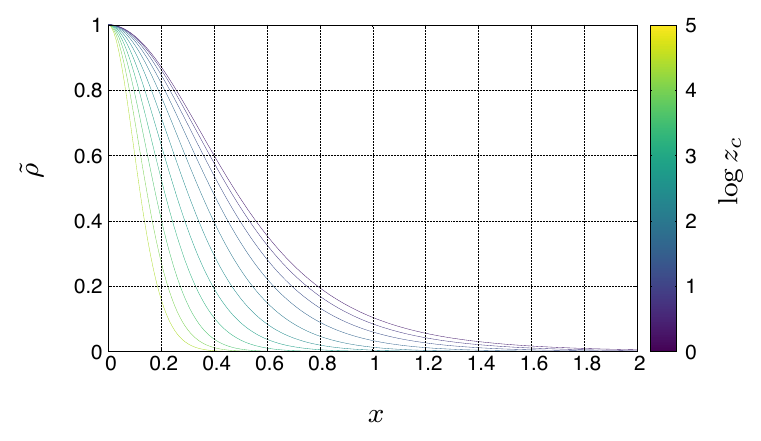}
        \subcaption{$z_c$ dependence for $\lambda = 1$.}
        \label{fig:dim5_beta_dep}
      \end{subfigure}
      \caption{
        The parameter dependence of the energy density for $D=5$.}
      \label{fig:dim5_density} 
    \end{figure}

\begin{figure}
      \begin{subfigure}[]{0.5\linewidth}
        \centering
        \hspace{-20pt}
        \includegraphics[width = 8cm]{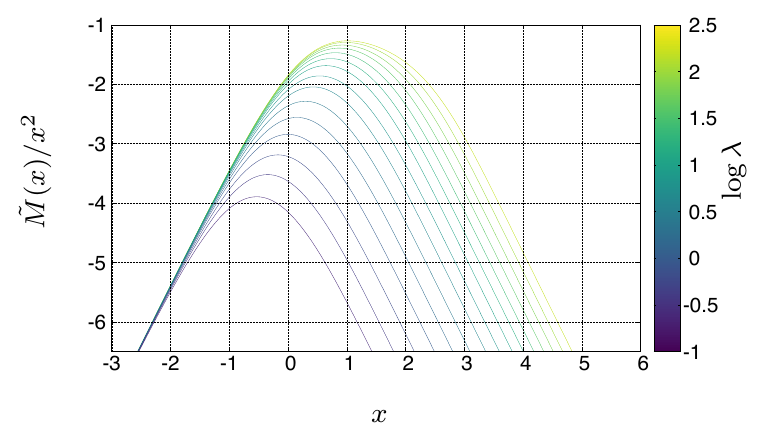}
        \subcaption{$\lambda$ dependence for $z_c = 1$.}
        \label{fig:dim5_lambda_dep_com}
      \end{subfigure}
      \begin{subfigure}[]{0.5\linewidth}
        \centering
        \includegraphics[width = 8cm]{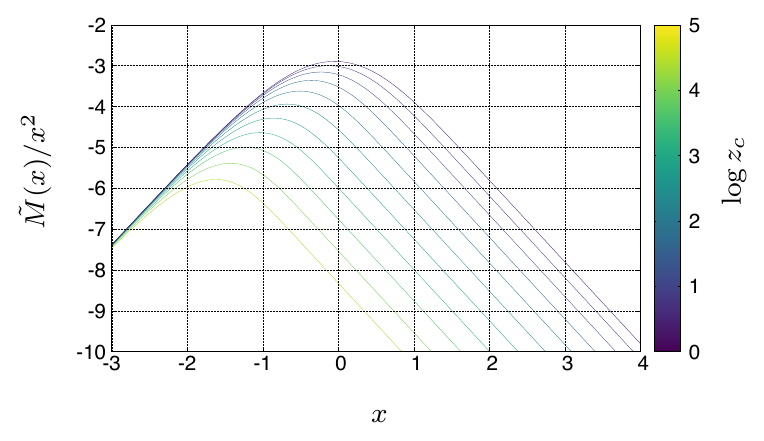}
        \subcaption{$z_c$ dependence for $\lambda = 1$.}
        \label{fig:dim5_beta_dep_com}
      \end{subfigure}
      \caption{
        The parameter dependence of the compactness parameter for $D=5$.
        It behaves as $\tilde{M}(x)/x^{2} \sim x^2$ for $x \ll 1$ and $\tilde{M}(x)/x^2 \sim x^{-2}$ for $x \gg \lambda$.}
      \label{fig:dim5_compact} 
    \end{figure}

    \figref{fig:dim5_density} shows the energy density profile as a function of $x$ for each parameter set $\pqty{z_c, \lambda}$ in the $D=5$ case. 
    \figref{fig:dim5_lambda_dep} indicates that the system becomes more compact as $\lambda$ decreases. 
    This is because the AdS potential works as a wall for the particles and the AdS radius $\lambda$ characterizes the size of the system.
    \figref{fig:dim5_beta_dep} indicates that the system becomes more compact as $z_c$ increases. 
    As mentioned above, since the parameter $z_c$ characterizes the central redshift factor, it can be regarded as an indicator of the significance of the relativistic gravitational effect in the system.
    That is, in the low temperature limit $z_c\to\infty$, the Newtonian limit can be realized.
    Actually, the compactness parameter, which is also an indicator of the relativistic effect, gets smaller as $z_c$ increases 
    as is shown in \figref{fig:dim5_beta_dep_com}, 
    and the relativistic effect of the system gets weaker. 

\begin{figure}
      \begin{subfigure}[]{0.5\linewidth}
        \centering
        \hspace{-20pt}
        \includegraphics[width = 8cm]{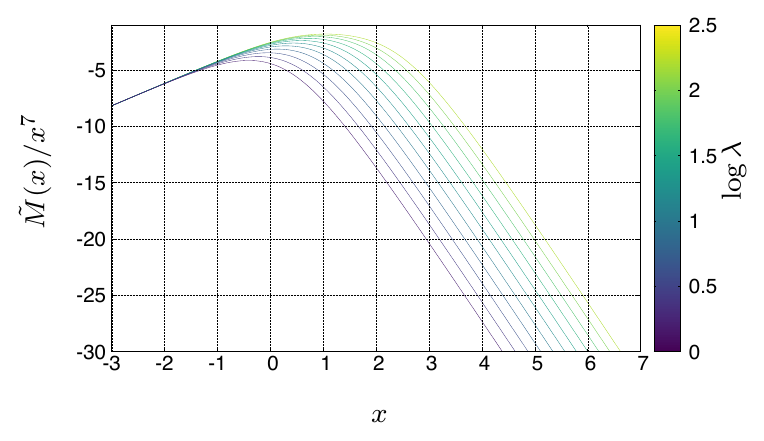}
        \subcaption{$\lambda$ dependence for $z_c = 1$.}
        \label{fig:dim10_lambda_dep_com}
      \end{subfigure}
      \begin{subfigure}[]{0.5\linewidth}
        \centering
        \includegraphics[width = 8cm]{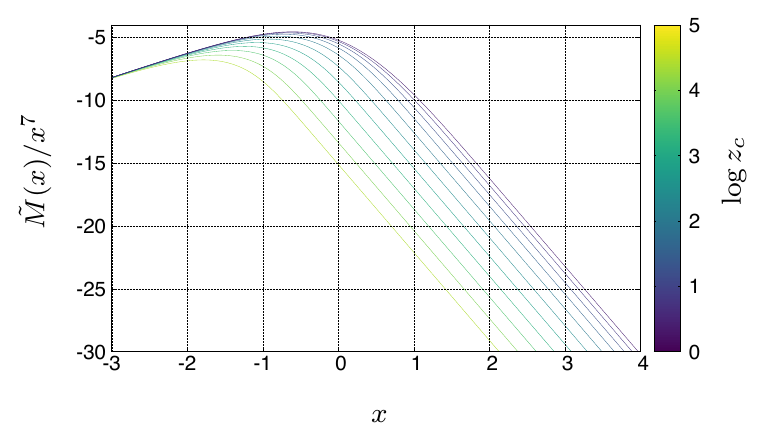}
        \subcaption{$z_c$ dependence for $\lambda = 1$.}
        \label{fig:dim10_beta_dep_com}
      \end{subfigure}
      \caption{
        The parameter dependence of the compactness parameter for $D=10$.
        It behaves as $\fcn{\tilde{M}}{x}/x^{7} \sim x^2$ for $x \ll 1$ and $\fcn{\tilde{M}}{x}/x^7 \sim x^{-7}$ for $x \gg \lambda$.}
      \label{fig:dim10_compact} 
    \end{figure}

    \figref{fig:dim5_compact} also shows that the total mass is constant for $x\gg\lambda$ because $\fcn{\tilde{M}}{x}/x^2 \sim x^{-2}$.
    In a $D$-dimensional spacetime, the compactness parameter behaves as $\fcn{\tilde{M}}{x}/x^{D-3} \sim x^{D-3}$ and 
    the total mass becomes constant (see \figref{fig:dim10_compact} for the $D=10$ case).
    These results reflect the confined structure of the self-gravitating system in the asymptotically AdS spacetime due to the AdS potential.
    Actually, both the total mass and the total particle number diverge for $\Lambda = 0$ without an artificial wall, 
    which is shown numerically in Ref.~\cite{Asami_2021} for the $D=4$ case.
    For $x\ll1$, $\dv*{\tilde{\rho}}{x} \sim \dv*{y}{x} = 0$ and $\fcn{\tilde{M}}{x}\sim x^{D-1}$ for $x \ll 1$, so that the compactness parameter behaves as $\fcn{\tilde{M}}{x}/x^{2} \sim x^2$ independently of the number of dimensions.
    
      
    \subsection{Turning point methods and parameter sets}
    \label{sec:method}
    Since the Einstein's equations \eqref{eq:ein_norm} and \eqref{eq:ein_mass} are uniquely determined by 
    the parameters $\pqty{z_c,\lambda}$, 
    they give a sequence of the solutions as a two parameter family of $\pqty{z_c,\lambda}$.
    In order to investigate the thermodynamical stability of each solution, 
    we regard $(\tilde{E},\tilde{L})$ as a parameter set specifying an equilibrium solution.
    Defining $\gamma := \log \tilde{L}$ and considering solutions with a fixed value of $\gamma$, 
    these solutions reduce to a one-parameter family.
    Introducing the sharpness parameter $\sigma := -\log\tilde{\rho}(\lambda)$ as the remaining one parameter, 
    the physical quantities listed in Eq.~\eqref{phys_quat} can be expressed by $\tilde{E} = \tilde{E}(\sigma)$, 
    $\tilde{L} = \tilde{L}(\sigma)$ and $\tilde{r} = \tilde{r}(\sigma)$.
    By using these values, we can define 
    \begin{align}
      (\bar{E},\bar{\beta})=\eval{\pqty{\frac{\lambda^{D-3}\fcn{\tilde{E}}{x}}{\fcn{\tilde{N}}{x}^2},\frac{\beta\fcn{\tilde{N}}{x}}{\lambda^{D-3}}}}_{x\to\infty}
      \label{eq:parameters_th}
    \end{align}
    for each value of $\gamma$, and of course, $\bar E$ and $\bar\beta$ are parametrized by $\sigma$ i.e., 
    $(\bar{E},\bar{\beta}) = (\bar{E}(\sigma),\bar{\beta}(\sigma))$.
    For a fixed value of $\gamma$, the series of equilibria draws a one-parameter curve in the $(\bar{E},\bar{\beta})$ space.
  
    The series of equilibria \eqref{eq:parameters_th} specifies the thermodynamical stability of the self-gravitating system, 
    and the stability can be checked through the turning point method\cite{Poincare1885,katz1978,Katz1979}.
    According to the turning point method, 
    the stability of the system with a fixed total particle number can change at points where the curve is vertical in the $(\bar{E},\bar{\beta})$ space.
    These points are called turning points. 
    The turning point method also says that the series of equilibria gets more unstable if it passes the turning point in the counter-clockwise direction.
  
    \subsection{Double spiral structure and the catastrophic instabilities}
    \label{sec:structure}
    \begin{figure}
        \begin{subfigure}[]{0.5\linewidth}
          \centering
          \hspace{-20pt}
          \includegraphics[width = 8cm]{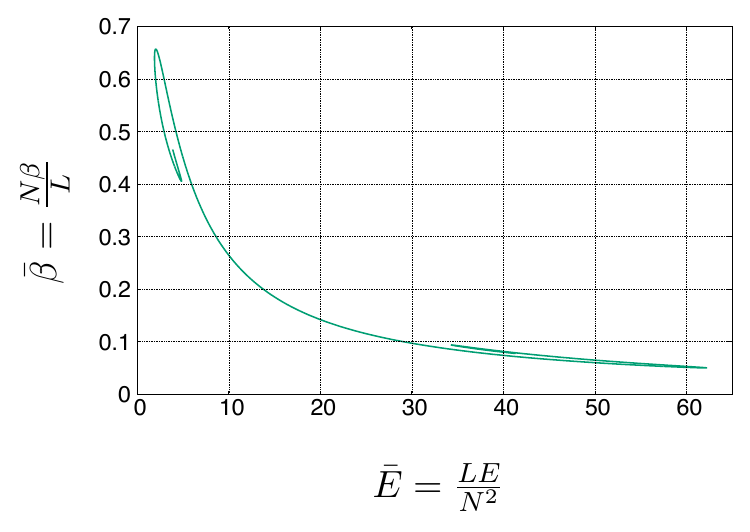}
          \subcaption{Inverse temperature: $\bar{\beta}=\bar{\beta}(\bar{E})$.}
          \label{fig:TypicalCurve}
        \end{subfigure}
        \begin{subfigure}[]{0.5\linewidth}
          \centering
          \includegraphics[width = 8cm]{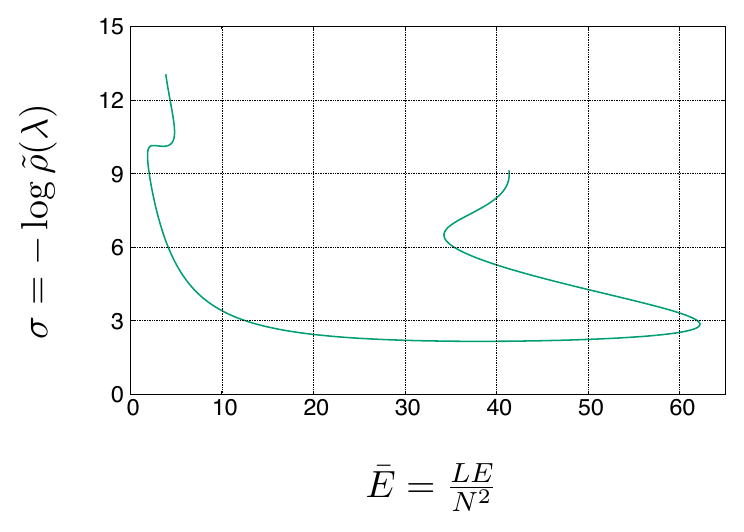}
          \subcaption{Sharpness parameter: $\sigma=\sigma(\bar{E})$.}
          \label{fig:DensityCurve}
        \end{subfigure}
      \caption{
        \figref{fig:TypicalCurve} shows the series of equilibria and \figref{fig:DensityCurve} shows the sharpness parameter $\sigma(\bar{E})$
        with $\gamma = 2.7$ in the $D=4$ case. 
        We can see that the curve has a double spiral structure and $\sigma$ increases from the middle to the both sides.
        The parameter $\bar E$ oscillates along the curve towards the both ends, and the curve forms spirals. 
      }
      \label{fig:Tpcl_curve_4}
    \end{figure}
    \figref{fig:Tpcl_curve_4} shows the inverse temperature $\bar{\beta}$ as a function of the gravothermal energy $\bar{E}$ of 
    the series of equilibria with $\gamma = 2.7$ for the 4-dimensional case, and the sharpness parameter $\sigma(\bar{E})$ is also plotted.
    The curve has a double spiral structure and 
    we call the spiral in the high energy region ``hot spiral'' and one in the low energy region ``cold spiral'', respectively.
    It would be noted that 
    a similar structure also appears in the case of self-gravitating particles inside an artificial wall with a vanishing cosmological constant~\cite{Alberti:2019xaj,Chavanis2020}. 
    The structure implies that the system has two catastrophic instabilities due to the negative specific heat of gravity.
    Since each spiral is spiraling in the counter-clockwise direction towards the center,
    the series of equilibria gets more unstable every time it passes a turning point.
    As is shown in \figref{fig:DensityCurve}, 
    the sharpness parameter $\sigma$ increases towards the center of each spiral and the particle distribution gets sharper.
    The two spirals are characterized by the total mass compactness $M/L$.
    Since we can rewrite $M/L = N/L(1+N\bar{E}/L) = \e{-\gamma}(1+\e{-\gamma}\bar{E})$,
    given a fixed value of $\gamma$, we can say the system is more compact for a larger value of $\bar{E}$.
    Based on this fact, we can understand the physical meaning of the instability on each spiral.
    On the cold spiral, the compactness of the system is relatively small.
    That implies particles can take energy away from the central region and the system causes a gravothermal catastrophe corresponding to the Newtonian one.
    By contrast, the compactness of the system is relatively large on the hot spiral.
    This behavior is peculiar to the relativistic system and it indicates the hot spiral reflects the strong non-linear gravity.
    If the gravity is sufficiently strong, 
    the system cannot support the configuration of the particles and experiences a catastrophic gravitational collapse.
  
    The double spiral structure ensures the existence of another instability.
    As is clear from the figures, the existence of the two spirals delimits the possible parameter region of $\bar E$ in a finite region.
    The fact tells us that no equilibrium solutions exist outside the region, 
    which is called the ``strong instability'' in contrast with the weak instability associated with a turning point. 
    Therefore, both cold and hot spirals ensure two kinds of catastrophic instability.
  
    \subsection{Configuration of curves and parameter dependence for $4\le D\le 10$}
    \label{sec:low_dim}
\begin{figure}
        \begin{subfigure}[]{0.5\linewidth}
          \centering
          \hspace*{-20pt}
          \includegraphics[width = 8cm]{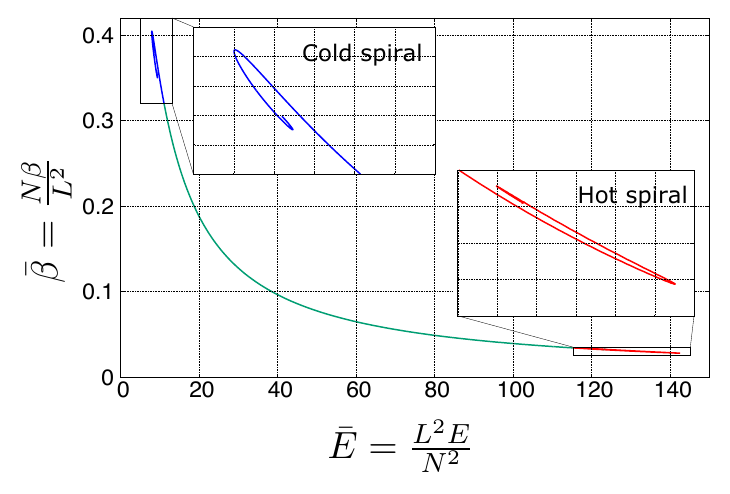}
          \subcaption{Inverse temperature: $\bar{\beta}=\bar{\beta}(\bar{E})$.}
          \label{fig:TypicalCurve_dim5}
        \end{subfigure}
        \begin{subfigure}[]{0.5\linewidth}
          \centering
          \includegraphics[width = 8cm]{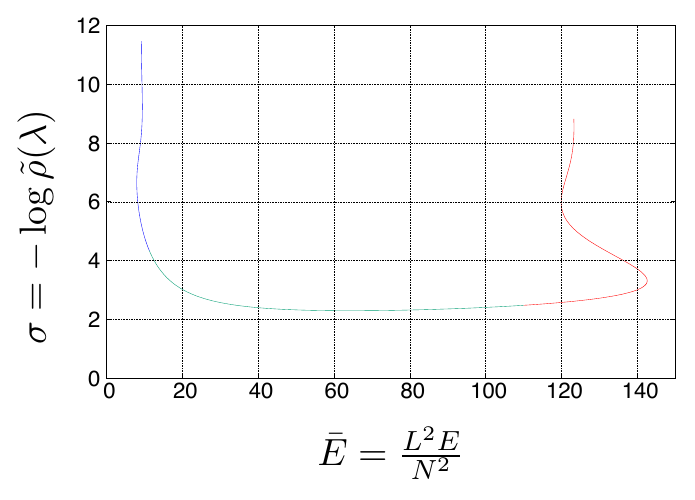}
          \subcaption{Sharpness parameter: $\sigma=\sigma(\bar{E})$.}
          \label{fig:DensityCurve_dim5}
        \end{subfigure}
      \caption{
        The inverse temperature and the sharpness parameter for a series of equilibrium states in the 5-dimensional case.
        We set $\gamma=2.65$ and the curve has a double spiral structure as in the 4-dimensional case.
      }
      \label{fig:Tpcl_curve_5}
    \end{figure}

    In the case of $4\le D\le 10$, the curve for thermal equilibrium states also has a double spiral structure.
    \figref{fig:Tpcl_curve_5} shows the inverse temperature $\bar{\beta}(\bar{E})$ 
    and the sharpness parameter $\sigma(\bar{E})$ for $\gamma=2.65$ in the 5-dimensional case.
    The embedded figures in \figref{fig:TypicalCurve_dim5} are enlarged figures of the cold and hots spirals. 
    We can see that both of spirals are counter-clockwise towards the center.
    Therefore, as in the 4-dimensional case, the system gets more unstable as the state approaches the center of each spiral.
  The parameter $\gamma = \log\tilde{L}^{D-3}$ denotes the rest mass compactness of the system.
  In the dilute limit, i.e. $\gamma\to\infty$, the hot spiral moves to the right infinitely, and the series of equilibria has only the cold spiral.
  In the dilute limit, the gravitational interaction becomes relatively less effective, and the Newtonian approximation would be valid. 
  Then only the Newtonian catastrophic instability is realized on the cold spiral.
  
\begin{figure}
      \begin{subfigure}[]{0.5\linewidth}
        \centering
        \hspace{-20pt}
        \includegraphics[width = 8cm]{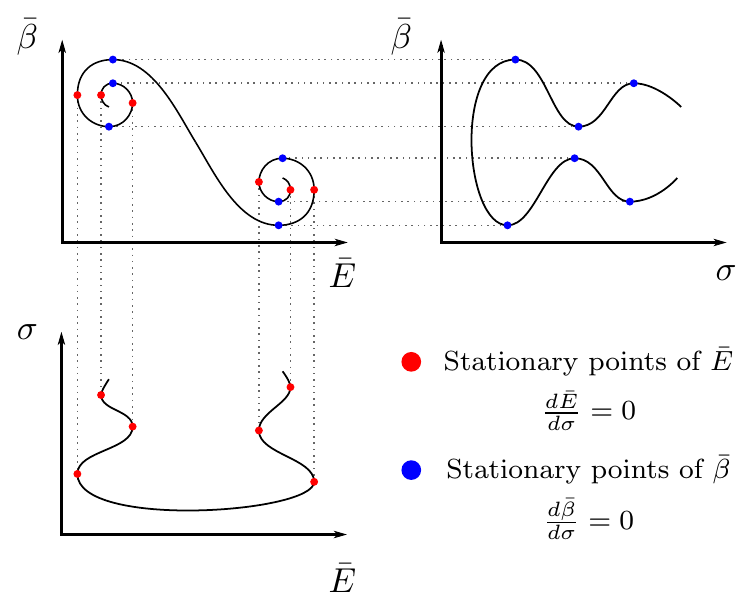}
        \subcaption{Large $\gamma$.}
        \label{fig:Schematic_pic_double_sp}
      \end{subfigure}
      \begin{subfigure}[]{0.5\linewidth}
        \centering
        \includegraphics[width = 8cm]{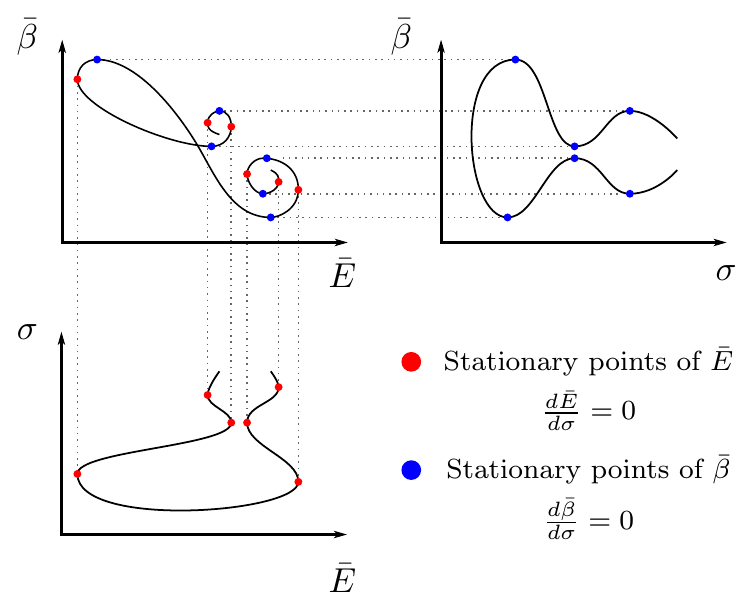}
        \subcaption{Small $\gamma$.}
        \label{fig:Schematic_pic_move}
      \end{subfigure}
      \caption{
        Schematic figures for the transition of configuration of the double spiral structure.
        The series of equilibria has no intersection in the $(\bar{E},\bar{\beta})$ plane for relatively large $\gamma$.
        As $\gamma$ decreases, the cold spiral moves to the right in the $(\bar{E},\bar{\beta})$ plane,
        and it intersects with the main part of the curve between the cold and hot spirals.
      }
      \label{fig:Schematic_picture_1}
    \end{figure}

\begin{figure}
    \begin{subfigure}[]{0.5\linewidth}
      \centering
      \hspace{-20pt}
      \includegraphics[width = 8cm]{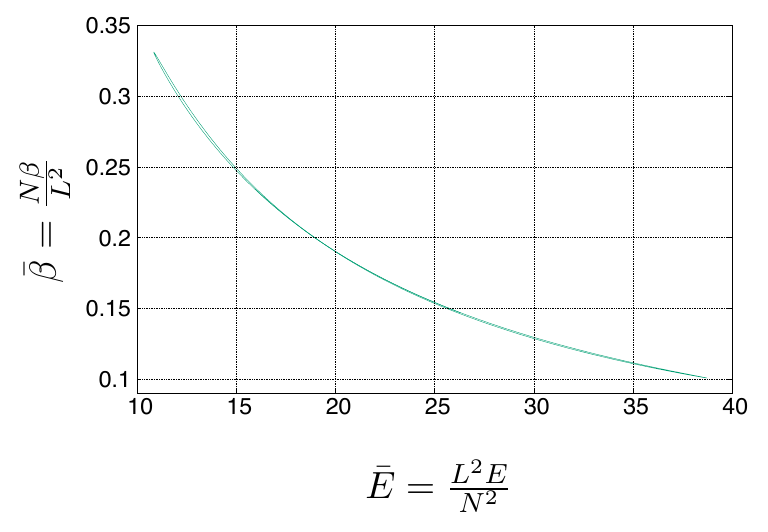}
      \subcaption{Inverse temperature: $\bar{\beta}=\bar{\beta}(\bar{E})$.}
      \label{fig:merge_point_curve_dim5}
    \end{subfigure}
    \begin{subfigure}[]{0.5\linewidth}
      \centering
      \includegraphics[width = 8cm]{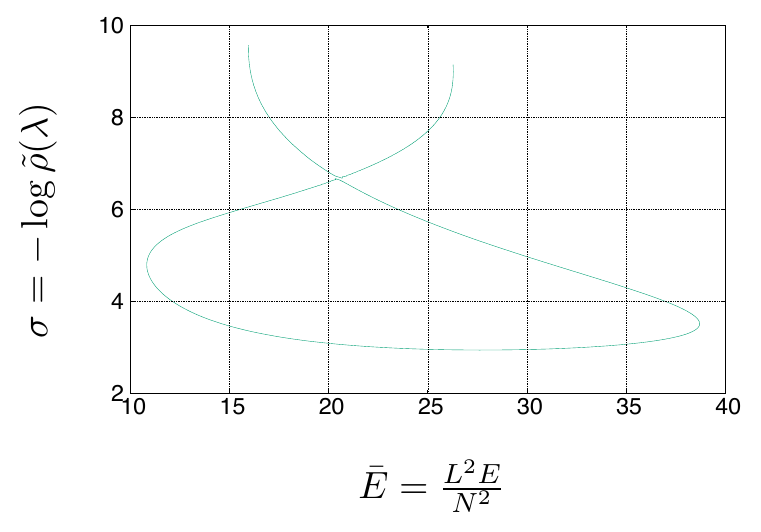}
      \subcaption{Sharpness parameter: $\sigma=\sigma(\bar{E})$.}
      \label{fig:merge_point_density_dim5}
    \end{subfigure}
    \caption{
      The curves of equilibrium states in the $(\bar{E},\bar{\beta})$ and $(\bar{E},\sigma)$ planes for $\gamma\simeq2.2525$.
      The cold and hot spirals merge at this point.
    }
    \label{fig:Tpcl_curve_5_merge}
  \end{figure}

\begin{figure}
      \begin{subfigure}[]{0.5\linewidth}
        \centering
        \hspace{-20pt}
        \includegraphics[width = 8cm]{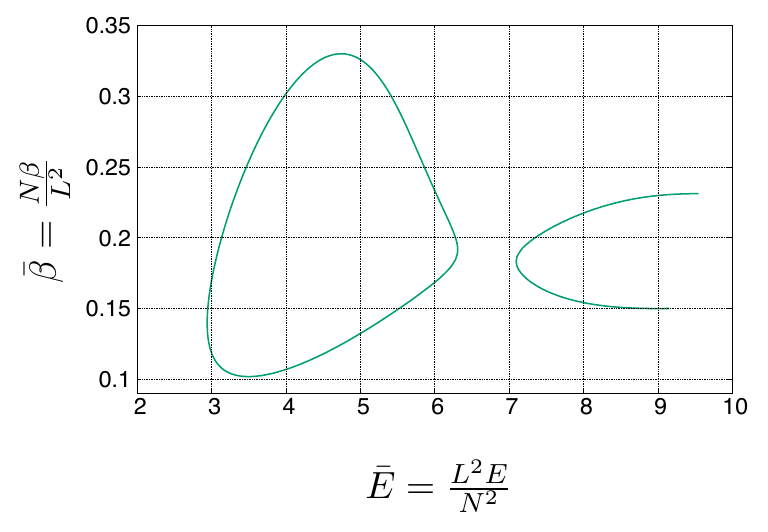}
        \subcaption{Inverse temperature: $\bar{\beta}=\bar{\beta}(\bar{E})$.}
        \label{fig:separated_curve_dim5}
      \end{subfigure}
      \begin{subfigure}[]{0.5\linewidth}
        \centering
        \includegraphics[width = 8cm]{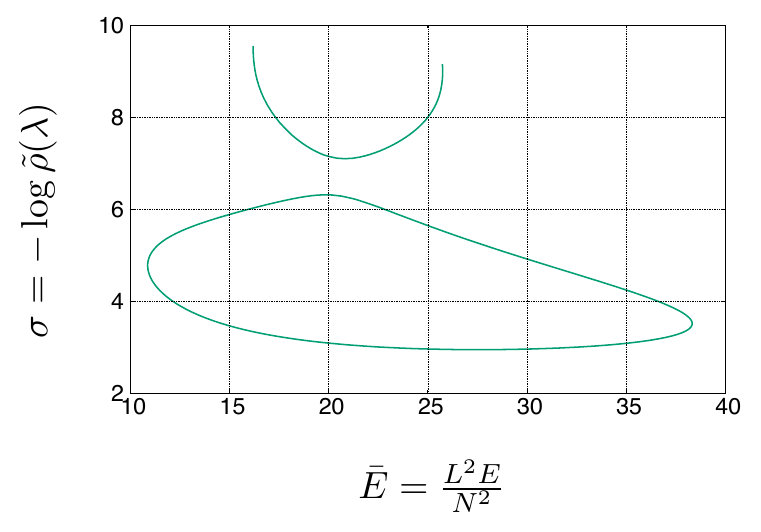}
        \subcaption{Sharpness parameter: $\sigma=\sigma(\bar{E})$.}
        \label{fig:separated_density_dim5}
      \end{subfigure}
    \caption{
      The curves of equilibrium states in the $(\bar{E},\bar{\beta})$ and $(\bar{E},\sigma)$ planes for $\gamma\simeq2.25$.
      In this parameter region, the series of equilibria separate into two pieces and two curves exist in $(\bar{E},\sigma)$ and $(\sigma,\bar{\beta})$ planes. 
    }
    \label{fig:Tpcl_curve_5_break}
  \end{figure}

\begin{figure}
    \begin{subfigure}[]{0.5\linewidth}
      \centering
      \hspace{-20pt}
      \includegraphics[width = 8cm]{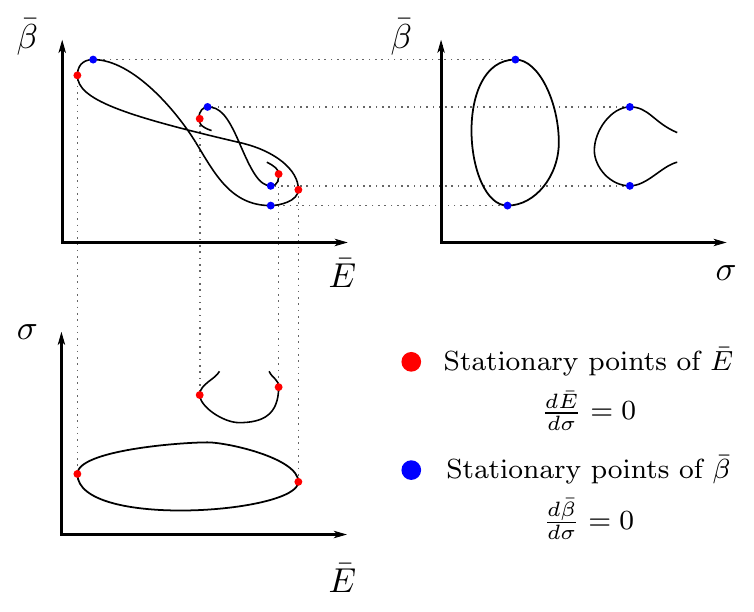}
      \subcaption{Separated sequences.}
      \label{fig:Schematic_pic_double_separated}
    \end{subfigure}
    \begin{subfigure}[]{0.5\linewidth}
      \centering
      \includegraphics[width = 8cm]{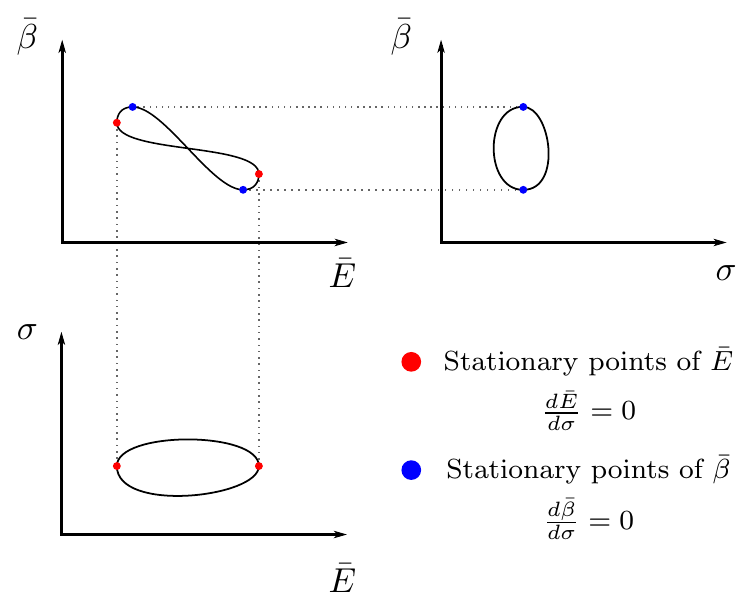}
      \subcaption{A loop structure.}
      \label{fig:Schematic_pic_loop}
    \end{subfigure}
    \caption{
      Schematic figures around the loop point $\gamma_\ell$.
      The sequence without loop structure shrinks and vanishes at this point.
      After the loop point, the other sequence shrinks until it reaches to the vanishing point $\gamma_\mathrm{v}$.
    }
    \label{fig:Schematic_picture_2}
  \end{figure}

  Let us consider how the configuration of the curve changes as $\gamma$ decreases.
  Two spirals approach each other as $\gamma$ decreases (schematic figures are shown in \figref{fig:Schematic_picture_1}), 
  and the cold spiral intersects with the main part of the curve between the two spirals. 
  In this process, a set of stationary points for $\bar E$ and $\bar \beta$ on each spiral approaches a common single point,
  and they merge at the ``merge point'' $\gamma_\mathrm{m}\simeq2.2525$.
  \begin{figure}
      \centering
        \includegraphics[width = 68mm]{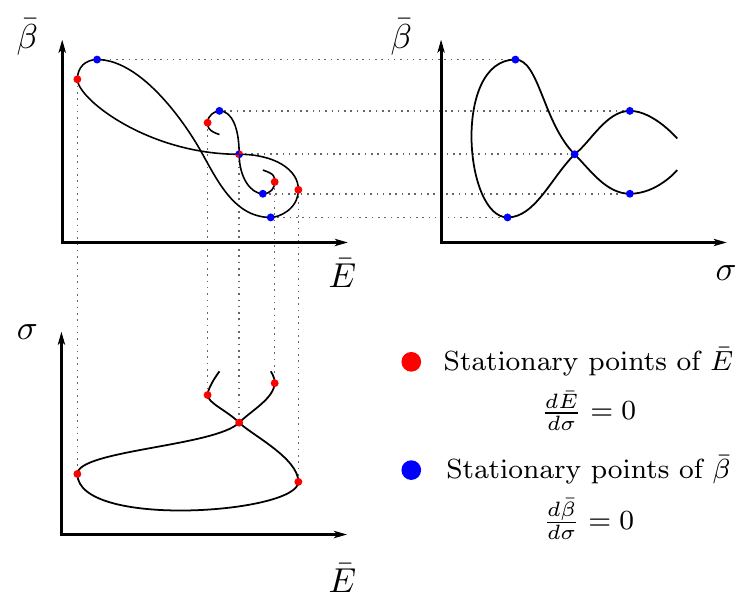}
        \caption{
          Schematic figures for the series of equilibria at the merge point.
          At the merge point, a set of stationary points for $\bar E$ and $\bar \beta$ on each spiral contract to a common single point. 
          The numerical result corresponding to the merge point is shown in \figref{fig:Tpcl_curve_5_merge}.
        }
        \label{fig:Schematic_picture_merge}
  \end{figure}
  \figref{fig:Tpcl_curve_5_merge} shows the configuration of the curves in the $(\bar{E},\bar{\beta})$ and $(\bar{E},\sigma)$ planes for $\gamma=2.2525$.
  For a value of $\gamma$ smaller than $\gamma_{\rm m}$, the curve of equilibria reconnects and separate into two sequences as shown in \figref{fig:Tpcl_curve_5_break}.
  We can see that one of sequences forms a loop and the other does not.
  As $\gamma$ gets even smaller, the sequence without a loop vanishes and the series of equilibria consists of only one loop (see \figref{fig:Schematic_picture_2}).
  We call this point the ``loop point'' and the value of $\gamma$ is $\gamma_\mathrm{\ell}\simeq2.2343$.
  \begin{figure}
      \begin{subfigure}[]{0.5\linewidth}
        \centering
        \hspace{-20pt}
        \includegraphics[width = 8cm]{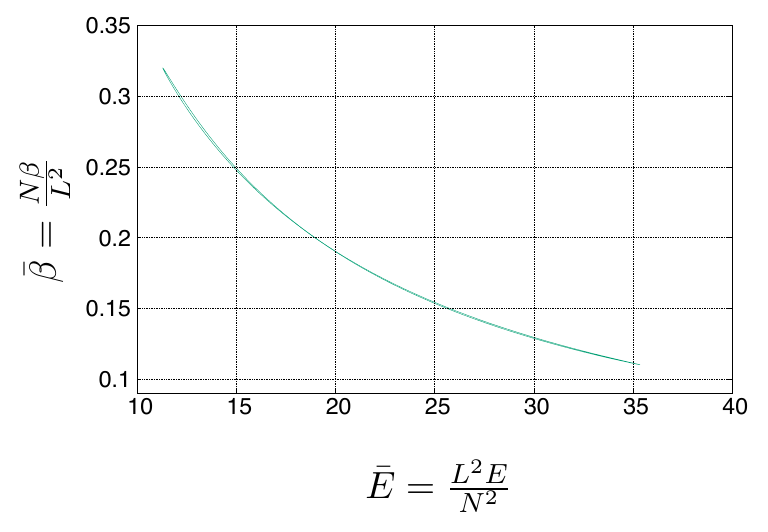}
        \subcaption{Inverse temperature: $\bar{\beta}=\bar{\beta}(\bar{E})$.}
        \label{fig:loop_point_curve_dim5}
      \end{subfigure}
      \begin{subfigure}[]{0.5\linewidth}
        \centering
        \includegraphics[width = 8cm]{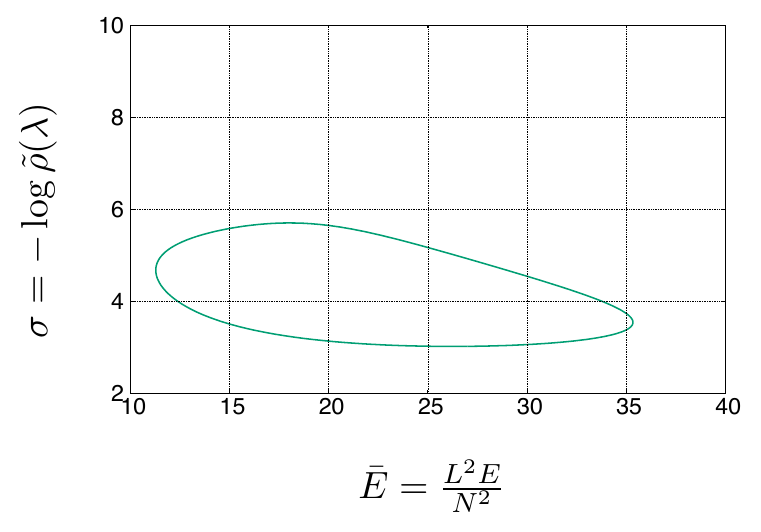}
        \subcaption{Sharpness parameter: $\sigma=\sigma(\bar{E})$.}
        \label{fig:loop_point_density_dim5}
      \end{subfigure}
    \caption{
      The curves of equilibrium states in the $(\bar{E},\bar{\beta})$ and $(\bar{E},\sigma)$ planes for $\gamma\simeq2.23$.
      The series has only the sequence with a loop.
      }
  \label{fig:Tpcl_curve_5_loop}
  \end{figure}
  If we continue to make $\gamma$ smaller, the sequence with a loop shrinks and vanishes at the ``vanishing point'' $\gamma_\mathrm{v}\simeq2.1369$.
  At the vanishing point, the series of equilibria becomes just a point.
  Therefore the thermal equilibrium state exists only at a unique gravothermal energy $\bar{E}_\ast\simeq18.17$.
  For $\gamma>\gamma_\mathrm{v}$, the system has no thermal equilibrium states, and strongly unstable. 
  
  In the $4\le D\le 10$ case, the behaviors of the series of equilibria are qualitatively similar to the 5-dimensional case.
  The series of equilibria has a double spiral structure and three critical points of $\gamma$.
  Therefore, the system has weak and strong instabilities associated with the turning points on both of cold and hot spirals.
  In any case, the catastrophic instabilities are caused by the negative specific heat of gravity.
  
  \subsection{Configuration of curves and parameter dependence in $D\ge11$ case}
  \label{sec:high_dim}
\begin{figure}
      \begin{subfigure}[]{0.5\linewidth}
        \centering
        \hspace{-20pt}
        \includegraphics[width = 8cm]{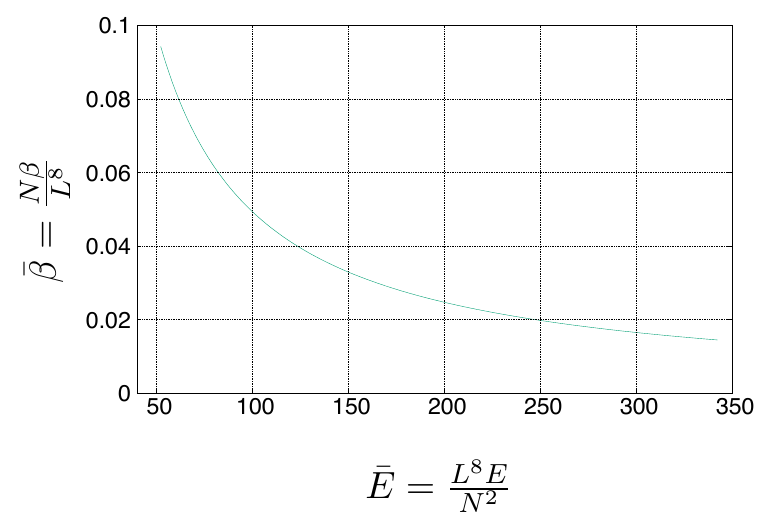}
        \subcaption{Inverse temperature: $\bar{\beta}=\bar{\beta}(\bar{E})$.}
        \label{fig:TypicalCurve_dim11}
      \end{subfigure}
      \begin{subfigure}[]{0.5\linewidth}
        \centering
        \includegraphics[width = 8cm]{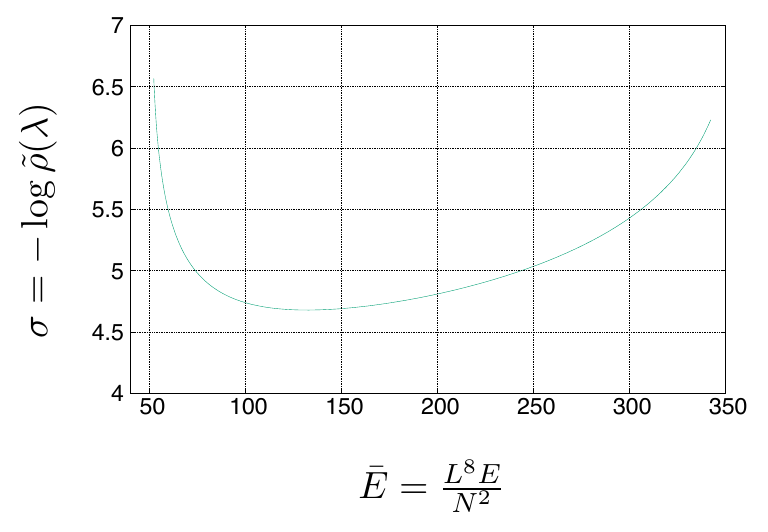}
        \subcaption{Sharpness parameter: $\sigma=\sigma(\bar{E})$.}
        \label{fig:DensityCurve_dim11}
      \end{subfigure}
    \caption{
      The inverse temperature and the sharpness parameter for a series of equilibrium states in the 11-dimensional case for $\gamma = 2.6$.
      The series does not have any spiral structures. 
      }
      \label{fig:Tpcl_curve_11}
  \end{figure}

  For $D\ge 11$, the behavior of the series of equilibria is dramatically different from the case for $4\le D\le 10$.
  \figref{fig:Tpcl_curve_11} shows the inverse temperature $\bar{\beta}(\bar{E})$ 
  and the sharpness parameter $\sigma(\bar{E})$ for $\gamma=2.6$ in the 11-dimensional case.
  We cannot see any spiral structures and stationary points disappear.
  This result implies the system does not have any weak instabilities due to the negative specific heat.
  In this case, the system has neither merge point nor loop point while it has a vanishing point around $\gamma_\mathrm{v}\simeq2.2271$.
  
  Similar critical behaviors have been reported for the systems confined by an artificial adiabatic wall~\cite{chavanis2002,Vaganov:2007at,Hammersley:2007ahw}.
  In our case, particles are confined by not an adiabatic wall but an AdS potential.
  We found that the critical dimension is exactly the same as the system confined by an adiabatic wall.
  The system still has the strong instabilities associated with the endpoints of the curve.
  This behavior is also qualitatively similar to the system with an adiabatic wall.

\section{Conclusion}
\label{sec:conclusion}
We have analyzed the structure and stability of the thermal equilibrium states of a self-gravitating system
in a $D$--dimensional asymptotically AdS spacetime. 
The thermal equilibrium can be realized because of the confinement of the AdS potential and we use the turning point method to investigate the stability.
We concluded that the properties of the stability are qualitatively similar to the relativistic system confined by an artificial wall~\cite{Alberti:2019xaj, Chavanis2020, Vaganov:2007at, Hammersley:2007ahw}. 

In a $D$--dimensional spacetime, the equilibria can be parameterized by two parameters contained in the solution for the Einstein's field equations.
Thus the series of equilibria draws a two-dimensional surface in the space of the rest mass compactness parameter $\gamma$,
the gravothermal energy $\bar{E}$ and the inverse temperature $\bar{\beta}$.
Given a fixed value of $\gamma$, the series reduces to a one-dimensional curve in the $(\bar{E},\bar{\beta})$ plane. 
As a result, the curve has a double spiral structure for $4\le D\le 10$ while it does not have any spirals for $D\ge 11$.

For $4\le D\le 10$, the system has two weak catastrophic instabilities corresponding to two spirals.
The cold spiral, which is located in the lower energy region, implies the gravothermal catastrophe associated with the fragmentation of the system.
The other spiral called a hot spiral, implies the catastrophic gravitational collapse. 
These properties indicate the existence of the strong instability since the allowed region of gravothermal energy is 
bounded by the turning points.
For $4\le D\le 10$, the configuration of the curve in the $(\bar{E},\bar{\beta})$ plane is dependent on the value of $\gamma$, 
and there are three critical points $\gamma_\mathrm{m}$, $\gamma_\mathrm{\ell}$ and $\gamma_\mathrm{v}$.
For $\gamma>\gamma_\mathrm{m}$, the series of equilibria has a double spiral structure.
Two spirals approach each other as $\gamma$ gets smaller. 
When the value of $\gamma$ gets smaller than $\gamma_\mathrm{m}$, the curve separates into two sequences of equilibria. 
One of sequences vanishes at $\gamma=\gamma_\mathrm{\ell}$.
Then, the series consists of only one loop for $\gamma_\mathrm{\ell}<\gamma<\gamma_\mathrm{v}$ and 
there is no equilibrium states for $\gamma>\gamma_\mathrm{v}$.

For $D\ge 11$, the system has no weak instabilities but strong instabilities corresponding to the endpoints of the curve in the $(\bar{E},\bar{\beta})$ plane.
Since the curve does not have any spiral structures, the system has neither the merge point nor the loop point.
The series of equilibria shrinks as $\gamma$ decreases and vanishes 
similarly to the cases for $4\le D\le 10$, and therefore it has a vanishing point $\gamma_\mathrm{v}$ as a critical point.

At the end, we conclude that the many-particle system has parameter regions with stable configuration. 
These configurations might well describe final fates of the turbulent instability in asymptotically AdS spacetimes. 
It should be noted that, however, the solutions of the many-particle system are not necessarily corresponding to stability islands discussed in Refs.~\cite{Choptuik_2018,Masachs_2019} because the system should be regarded as a macroscopic model after the non-linear turbulent phenomena. 
Apparently, we need further investigations to clarify implication of our results in the turbulent instability in asymptotically AdS spacetimes.

\section*{Acknowledgments}
This work was supported in part by JSPS KAKENHI Grant Nos.
JP19H01895 (CY), JP20H05850 (CY) and JP20H05853 (CY). 
    This work was also supported by JST SPRING, Grant Number JPMJSP2125 (HA). 
The author (HA) would like to take this opportunity to thank the 
“Interdisciplinary Frontier Next-Generation Researcher Program of the Tokai Higher Education and Research System.”

\appendix
  \def\thesection{Appendix \Alph{section}}

  \renewcommand{\theequation}{A.\arabic{equation}}
\setcounter{equation}{0}

\section{Derivation of the thermal equilibrium}
\label{sec:app_der}
In this appendix, from the extremal entropy condition~\eqref{eq:extremal}, 
we first derive the thermal equilibrium condition for the distribution function $f$ with the following general expression of the entropy:
\begin{align}
  \fcl{S}{\fcn{\mathcal{S}}{f}} := \int_\Sigma \dd{\Sigma_\mu}S^\mu\qc 
  \fcl{S^\mu}{\fcn{\mathcal{S}}{f}} = \int \dd{V_p}p^\mu\fcn{\mathcal{S}}{f}. 
  \label{eq:def_gen_ent}
\end{align}
Then we show that the condition for the BG entropy leads to the MJ distribution.

\subsection{Coordinate systems in the momentum space}
\label{sec:app_der_int}
First, let us introduce the local inertial frame whose metric components are related to $g_{\mu\nu}$ given in Eq.~\eqref{eq:metric} by 
\begin{align}
  \eta_{\hat{\mu}\hat{\nu}} = \mathcal{L}^{\mu}_{\ \hat{\mu}}\mathcal{L}^{\nu}_{\ \hat{\nu}}g_{\mu\nu},
\end{align}
where $\eta_{\hat{\mu}\hat{\nu}} = \mathrm{diag}\pqty{-1, 1, \cdots, 1}$ and 
\begin{align}
  \pqty{\mathcal{L}^{\mu}_{\ \hat{\mu}}} = \mathrm{diag}\pqty{\e{-\mu},\e{-\nu}, \frac{1}{r}, \cdots, \frac{1}{r\sin\theta_1\sin\theta_2\cdots\sin\theta_{D-3}}}.
\end{align}
In this frame, 
the invariant volume element in the momentum space is written as
\begin{align}
  dV_p 
  &= 2\delta(p^2+1)\theta(\hat{\veps})\dd[D]{p} \nt
  &= \frac{1}{\hat{\veps}} \dd{p}^{\hat{r}}\wedge\dd{p}^{\hat{\theta}_1}\wedge\cdots\wedge\dd{p}^{\hat{\theta}_{D-2}},
\end{align}
where $\hat{\veps} := -p_{\hat{t}}$.
The variation of the distribution function $f$ is taken as a function of $p_{\hat{\mu}}$, that is, $p_{\hat{\mu}}$ is fixed in the variation. 

For convenience, we also introduce the coordinates $(\veps,J,\psi_1,\psi_2,\cdots,\psi_{D-3})$ in the momentum space as follows:
\begin{align}
  p^{\hat{\theta}_1} &= \frac{\sqrt{J}}{r}\cos\psi_1, \nt
  p^{\hat{\theta}_2} &= \frac{\sqrt{J}}{r}\sin\psi_1\cos\psi_2,\nt
  &\hspace{5pt}\vdots \nt
  p^{\hat{\theta}_{D-3}} &= \frac{\sqrt{J}}{r}\sin\psi_1\sin\psi_2\cdots\sin\psi_{D-4}\cos\psi_{D-3}, \nt
  p^{\hat{\theta}_{D-2}} &= \frac{\sqrt{J}}{r}\sin\psi_1\sin\psi_2\cdots\sin\psi_{D-4}\sin\psi_{D-3},
\end{align}
where $J$ is the square of the total angular momentum.
The volume element is written as
\begin{align}
  \dd{V_p} = \frac{J^\frac{D-4}{2}}{2r^{D-2}\hat\veps}\dd{p^{\hat r}}\wedge\dd{J}\wedge\dd{\Omega_{D-3}}
\end{align}
and the relevant domain for $p^{\hat r}$ and $J$ is given by $\Bqty{(p^{\hat r},J)|-\infty < p^{\hat r} < \infty, 0 < J < \infty}$.
Regarding $\hat\veps$ as an independent variable, the invariant volume element is given by
\begin{align}
  \dd{V_p} = \frac{J^{\frac{D-4}{2}}}{r^{D-2}}\pqty{{\hat\veps}^2-\pqty{1+\frac{J}{r^2}}}^{-\frac{1}{2}}\dd{\hat \veps}\wedge\dd{J}\wedge\dd{\Omega_{D-3}},
  \label{eq:vol_eps}
\end{align}
and the relevant region is given by $\Bqty{(\hat \veps,J)|1 < \hat \veps < \infty, 0 < J < r^2({\hat\veps}^2-1)}$ due to the on-shell condition, 
where we multiplied the factor two for the degeneracy of the sign of $p^{\hat r}$.

\subsection{Variation of physical quantities}
\label{sec:app_der_eqs}
Since the system is isolated, we fix the total mass and the total particle number.
For this purpose, we calculate the variation of such quantities and the entropy in $f\to f+\delta f$.
Hereafter, the subscript $E$ denotes the equilibrium state, such as $\mu_E,\ \nu_E$ and $\rho_E$, and we consider the perturbation around it,
that is, $\mu = \mu_E + \delta\mu\ etc.$ with respect to the perturbation $\delta f$.

The perturbation of $\nu$, say $\delta\nu$, is determined by Eq.~\eqref{eq:ein_1} through the energy density as follows:
  \begin{align}
    \dv{r}(r^{D-3}\e{-2\nu_E}\var{\nu}) = \frac{8\pi}{D-2}r^{D-2}\var{\rho},
  \end{align}
where the perturbation of the energy density is given by
  \begin{align}
    \delta\rho :
    = \fcl{\rho}{f+\delta f} - \fcl{\rho}{f}
    = \int\dd{V_p} \hat{\veps}^2\delta f.
  \end{align}
Therefore, the perturbation $\delta\nu$ is not independent of $\delta f$ through the Einstein's equations.
However, it is convenient 
to formally treat them as independent perturbations in the following calculations.

The perturbation of $\fcl{F}{f}$ defined by Eq.~\eqref{eq:def_charge} is 
\begin{align}
  \delta{F} :
  &= \fcl{F}{f+\delta f} - \fcl{F}{f} \nt
  &= \int\dd{\Gamma} \hat{\veps}\delta\pqty{\e{\nu}\mathcal{F}} \nt
  &= \int\dd{\Gamma} \e{\nu}\hat{\veps}\bqty{\pqty{\mathcal{F}\delta\nu+\pdv{\mathcal{F}}{f}\delta f}
  + \order{\delta{f}^2}},
\end{align}
where $\dd{\Gamma} := \dd[D-1]{x}\dd{V_p} = \dd{r}\wedge\pqty{r\dd{\theta_1}}\wedge\cdots\wedge\pqty{r\sin\theta_1\sin\theta_2\cdots\sin\theta_{D-3}\dd\theta_{D-2}}\wedge\dd{V_p}$ is the measure in the phase space.

\subsection{Derivation}
\label{sec:app_der_der}
Up to the first order, the equilibrium condition \eqref{eq:extremal} becomes
\begin{align}
  \int \dd{r} r^{D-2} \int \dd{V_{p}} \Bqty{\e{\nu} \hat{\veps}\pqty{\mathcal{S} \var{\nu}+\pdv{\mathcal{S}}{f} \var{f}}+\alpha \e{\nu} \hat{\veps}\pqty{f \var{\nu}+\var{f}}-\beta \hat{\veps}^2 \var{f}}=0.
  \label{eq:cond1}
\end{align}
Introducing 
\begin{align}
  s^\ast := \pdv{\mathcal{S}}{f}+\alpha-\beta\veps = \pdv{\mathcal{S}}{f}+\alpha-\hat{\beta}\hat{\veps}
\end{align} 
with $\hat{\beta} := \e{\mu}\beta$, thermal equilibrium condition \eqref{eq:cond1} is rewritten as
\begin{align}
  \int \dd{r} r^{D-2} \int \dd{V_{p}} \pqty{\mathcal{S} \var{\nu}+ s^\ast\var{f}+\alpha f\var{\nu}}\e{\nu} \hat{\veps}
  + \int \dd{r} r^{D-2}\beta\pqty{\e{\mu+\nu}-1}\int\dd{V_p}\hat{\veps}^2\var{f}
  =0.
  \label{eq:cond2}
\end{align}
The second term becomes
\begin{align}
    & \int \dd{r} r^{D-2}\beta\pqty{\e{\mu+\nu}-1}\int\dd{V_p}\hat{\veps}^2\var{f} \notag \\
  = & \int \dd{r} \beta\pqty{\e{\mu+\nu}-1}\frac{\pqty{D-2}}{8\pi}\dv{r}(r^{D-3}\e{-2\nu}\var{\nu}) \notag \\
  = & -\frac{\pqty{D-2}\beta}{8\pi}\int\dd{r}\pqty{\mu^\prime+\nu^\prime}\e{\mu-\nu}r^{D-3}\var{\nu}  \notag \\
  = & -\beta\int\dd{r}r^{D-2}\e{\mu+\nu}\pqty{\rho+p}\var{\nu},
\end{align}
where we used the conservation of the energy-momentum:
\begin{align}
  \mu^\prime + \nu^\prime = \frac{8\pi}{D-2}r\e{2\nu}\pqty{\rho+p}.
\end{align}
Then the condition \eqref{eq:cond2} becomes
\begin{align}
  \int \dd{r} r^{D-2} \int \dd{V_{p}} \pqty{H \var{\nu}+ s^\ast\var{f}}\e{\nu} \hat{\veps}
  -\beta\int\dd{r}r^{D-2}\e{\mu+\nu}\pqty{\rho+p}\var{\nu}
  =0,
  \label{eq:cond3}
\end{align}
where $H=\mathcal{S}+\alpha f$.

We use the identity for $G = \fcn{G}{\veps,J}$:
\begin{align}
  \int\dd{V_p}\eval{\pdv{G}{\veps}}_{J}\pqty{p_t}^i\pqty{p_r}^j 
  = -i\int\dd{V_p}G\pqty{p_t}^{i-1}\pqty{p_r}^j -\pqty{j-1}\e{-2\mu+2\nu}\int\dd{V_p}G\pqty{p_t}^{i+1}\pqty{p_r}^{j-2}.
  \label{eq:identity}
\end{align}
The proof of this identity is given in \ref{sec:proof}.
Substituting $(i,j) = (1,2)$ into Eq.~\eqref{eq:identity}, we obtain
\begin{align}
    \rho+p 
  = & -\int\dd{V_p}p_t p^t f + \int\dd{V_p}p_r p^r f \notag \\
  = & \e{-2\nu}\Bqty{\int\dd{V_p}\pqty{p_r}^2 f + \e{-2\mu+2\nu}\int\dd{V_p}\pqty{p_t}^2 f} \notag \\
  = & -\e{-2\nu} \int\dd{V_p}\pqty{p_t}\pqty{p_r}^2\pdv{f}{\veps}\eval_{J}\notag \\
  = & \e{\mu} \int\dd{V_p}\hat{\veps}\ p_r p^r \pdv{f}{\veps}\eval_{J}.
\end{align}
Therefore, 
\begin{align}
    \beta\pqty{\rho+p}
  = & \beta\e{\mu} \int\dd{V_p}\hat{\veps}\ p_r p^r \pdv{f}{\veps}\eval_{J} \notag \\
  = & \int\dd{V_p}p_r p^r\beta\veps\pdv{f}{\veps}\eval_{J}.
\end{align}
Since 
\begin{align}
  \pdv{\mathcal{S}}{\veps}\eval_{m,J} = \pdv{\mathcal{S}}{f}\pdv{f}{\veps}\eval_{J} 
  = \pqty{s^\ast-\alpha+\beta\veps}\pdv{f}{\veps}\eval_{J}, 
\end{align}
\begin{align}
    \beta\pqty{\rho+p}
  = & \int\dd{V_p}p_r p^r\pqty{\pdv{\mathcal{S}}{\veps}\eval_{J}+\alpha\pdv{f}{\veps}\eval_{m,J} -s^\ast\pdv{f}{\veps}\eval_{J} }\notag \\
  = & \e{-2\nu}\int\dd{V_p}\pqty{p_r}^2\pdv{H}{\veps}\eval_{J} -\int\dd{V_p}p_r p^rs^\ast\pdv{f}{\veps}\eval_{J}.
\end{align}
Substituting $(i,j) = (0,2)$ into Eq.~\eqref{eq:identity}, we obtain
\begin{align}
    \beta\pqty{\rho+p}
  = & \e{-\mu}\int\dd{V_p}\hat{\veps}H-\int\dd{V_p}p_r p^r s^\ast\pdv{f}{\veps}\eval_{J}.
\end{align}
Therefore, Eq.~\eqref{eq:cond3} becomes
\begin{align}
  0 
  &= \int \dd{r} r^{D-2} \bqty{\int \dd{V_{p}} \pqty{H \var{\nu}+ s^\ast\var{f}}\e{\nu} \hat{\veps}-\beta\e{\mu+\nu}\pqty{\rho+p}\var{\nu}} \notag \\
  &= \int \dd{r} r^{D-2} \int \dd{V_{p}} \pqty{\hat{\veps}\var{f} + p_r p^r\e{\mu}\pdv{f}{\veps}\eval_{J}\var{\nu}}\e{\nu}s^\ast. \\
\end{align}
Since $\var{f}$ and $\var{\nu}$ are dependent variations, above condition for an arbitrary variation leads to the condition $s^\ast = 0$, i.e.,
\begin{align}
  \pdv{\mathcal{S}}{f}+\alpha-\beta\veps = 0.
\end{align}
For the BG entropy, the condition becomes $-\log f+\alpha-\beta\veps = 0$, which leads to the MJ distribution function.

\subsection{Proof of Eq.~\eqref{eq:identity}}
\label{sec:proof}

Using the coordinates with Eq.~\eqref{eq:vol_eps}, the left-hand side of Eq.~\eqref{eq:identity} becomes 
\begin{align}
  \int\dd{V_p}\eval{\pdv{G}{\veps}}_{J}\pqty{p_t}^i\pqty{p_r}^j 
  &= \int\dd{\veps}\wedge\dd{J}\wedge\dd{\Omega_{D-3}}\frac{J^{\frac{D-4}{2}}}{r^{2(D-2)}}\pqty{\veps^2-\veps_0^2\pqty{1+\frac{J}{r^2}}}^{-\frac{1}{2}}\eval{\pdv{G}{\veps}}_{J}\pqty{p_t}^i\pqty{p_r}^j. 
\end{align}
Since 
\begin{align}
  &\eval{\pdv{}{\veps}}_{J}\pqty{\pqty{\veps^2-\veps_0^2\pqty{1+\frac{J}{r^2}}}^{-\frac{1}{2}}\pqty{p_t}^i\pqty{p_r}^j}\nt
  = &\e{j(-\mu+\nu)}\eval{\pdv{}{\veps}}_{J}\pqty{\pqty{-\veps}^{i}\pqty{\veps^2-\veps_0^2\pqty{1+\frac{J}{r^2}}}^{\frac{j-1}{2}}}\nt
  = &\e{j(-\mu+\nu)}\pqty{-i\pqty{-\veps}^{i-1}\pqty{\veps^2-\veps_0^2\pqty{1+\frac{J}{r^2}}}^{\frac{j-1}{2}}-\pqty{j-1}\pqty{-\veps}^{i+1}\pqty{\veps^2-\veps_0^2\pqty{1+\frac{J}{r^2}}}^{\frac{j-3}{2}}}\nt
  = &\pqty{\veps^2-\veps_0^2\pqty{1+\frac{J}{r^2}}}^{-\frac{1}{2}} \bqty{-i\pqty{p_t}^{i-1} \pqty{p_r}^j - \pqty{j-1} \e{-2\mu+2\nu} \pqty{p_t}^{i+1} \pqty{p_r}^{j-2}},
\end{align}
integrating by parts, we get Eq.~\eqref{eq:identity}.\ $\square$

  \renewcommand{\theequation}{B.\arabic{equation}}
  \setcounter{equation}{0}
  \section{The integration over the momentum space}
  \label{sec:app_int_mom}
  We derive the explicit expressions for the physical quantities by performing the integration over the momentum space.
  In order to simplify the integrations, we transform the integral variables as $J\to s := J/J_\mathrm{max}$, 
  where $J_\mathrm{max}:=r^2({\hat \veps}^2-1)$ is the upper bound for the integration of $J$.
  In the coordinate system $\pqty{\hat \veps,s,\psi_1,\psi_2,\cdots,\psi_{D-3}}$, the invariant volume element is rewritten as 
  \begin{align}
      dV_p = \frac{s^{\frac{D-4}{2}}}{\sqrt{1-s}}({\hat \veps}^2-1)^\frac{D-3}{2}\dd{\hat \veps}\wedge\dd{s}\wedge\dd{\Omega_{D-3}},
    \end{align}
  where $1 \leq \hat\veps < \infty$ and $0 \leq s \leq 1$.
  If the distribution function takes the form $f = \fcn{f}{\hat \veps,J}$, the energy density is given by
  \begin{align}
    \rho(r)
    &= \int dV_p\ \hat\veps^2 f(\hat \veps,J)\nt
    &= S_{D-2}\int_{1}^\infty \dd{\hat \veps}\int_0^1 \dd{s}\frac{s^{\frac{D-4}{2}}(\hat \veps^2-1)^\frac{D-3}{2}}{\sqrt{1-s}}\hat \veps^2f,
  \end{align}
  and the pressure becomes
  \begin{align}
    p(r)
    &= \int dV_p\ (p^{\hat r})^2 f(\hat \veps,J) \nt
    &= S_{D-2}\int_{1}^\infty \dd{\hat \veps}\int_{0}^{1} \dd{s}s^{\frac{D-4}{2}}\sqrt{1-s}(\hat \veps^2-1)^\frac{D-1}{2}f.
  \end{align}
  For the MJ distribution function $f(\hat \veps)=\exp(\alpha-z\hat{\veps})$, by using
  \begin{align}
    \int_0^1\dd{s}\frac{s^{\frac{D-4}{2}}}{\sqrt{1-s}} &= \frac{\sqrt{\pi}\Gamma(\frac{D-2}{2})}{\Gamma(\frac{D-1}{2})}, \nt
    \int_0^1\dd{s}s^{\frac{D-4}{2}}\sqrt{1-s} &= \frac{\sqrt{\pi}\Gamma(\frac{D-2}{2})}{2\Gamma(\frac{D+1}{2})},
  \end{align}
  we obtain
    \begin{subequations}
      \begin{align}
        \rho(r) &= S_{D-1}\e{\alpha}\int_{1}^\infty \dd{\hat \veps}(\hat \veps^2-1)^\frac{D-3}{2}\hat\veps^2\e{-z\hat \veps}, \\
        p(r)    &= \frac{S_{D-1}}{D-1}\e{\alpha}\int_{1}^\infty \dd{\hat \veps}(\hat \veps^2-1)^\frac{D-1}{2}\e{-z\hat \veps}.
      \end{align}
    \end{subequations}
    Introducing the modified Bessel function of the second kind $\fcn{K_n}{z}$:
    \begin{align}
      \fcn{K_n}{z} = \frac{\sqrt{\pi}\pqty{\frac{z}{2}}^n}{\Gamma\pqty{n+\frac{1}{2}}}\int_{1}^\infty \dd{t}(t^2-1)^{n-\frac{1}{2}}\e{-zt},
    \end{align}
    the energy density and the pressure become
    \begin{subequations}
      \begin{align}
        \rho(r)
        &= 2\e{\alpha}\pqty{\frac{2\pi}{z}}^{\frac{D-2}{2}}\pqty{\frac{D-1}{z}K_{\frac{D}{2}}(z)+K_{\frac{D-2}{2}}(z)}, \\
        p(r)
        &= \frac{\e{\alpha}}{\pi}\pqty{\frac{2\pi}{z}}^{\frac{D}{2}}K_{\frac{D}{2}}(z).
      \end{align}
    \end{subequations}
    For the particle number, we can rewrite its density as
    \begin{align}
      n(r):
      = &\e{-\mu+\nu}\int \dd{V_p} \veps \e{\alpha-\beta\veps} \nt
      = &\e{\alpha+\nu}\frac{2\pi^{\frac{D-1}{2}}}{\Gamma(\frac{D-1}{2})}\int_{1}^\infty\dd{\hat{\veps}}({\hat{\veps}}^2-1)^\frac{D-3}{2}{\hat{\veps}}\e{-z{\hat{\veps}}} \nt
      = &2\e{\alpha+\nu}\pqty{\frac{2\pi}{z}}^{\frac{D-2}{2}}K_{\frac{D}{2}}(z),
    \end{align}
    where we used
    \begin{align}
      \int_1^\infty \dd{{\hat{\veps}}} \pqty{{\hat{\veps}}^2-1}^{n-\frac{1}{2}}{\hat{\veps}}\e{-z{\hat{\veps}}}
      = &\frac{1}{2n+1}\int_1^\infty \dd{{\hat{\veps}}} \dv{{\hat{\veps}}}\pqty{\pqty{{\hat{\veps}}^2-1}^{n+\frac{1}{2}}}\e{-z{\hat{\veps}}} \nt
      = &\frac{1}{2n+1}\bqty{\pqty{{\hat{\veps}}^2-1}^{n+\frac{1}{2}}\e{-z{\hat{\veps}}}}_{1}^{\infty}+\frac{z}{2n+1}\int_1^\infty \dd{{\hat{\veps}}} \pqty{{\hat{\veps}}^2-1}^{n+\frac{1}{2}}\e{-z{\hat{\veps}}} \nt
      = &\frac{\Gamma(n+\frac{1}{2})}{\sqrt{\pi}\pqty{\frac{z}{2}}^{n}}K_{n+1}(z).
    \end{align}

\bibliographystyle{unsrt}
  \bibliography{citation.bib}

\end{document}